\documentclass[aps,superscriptaddress,showpacs,nofootinbib,twocolumn]{revtex4}
\usepackage{graphics}
\usepackage{bm}
\usepackage{graphicx}
\usepackage{amsmath}
\usepackage{amssymb}
\usepackage{float}
\usepackage{caption}
\usepackage{subfigure}

\newcommand{\be}{\begin{equation}}
\newcommand{\ee}{\end{equation}}
\newcommand{\bea}{\begin{eqnarray}}
\newcommand{\eea}{\end{eqnarray}}
\begin{document}

\title{High dimensional chaotic systems which behave like random walks in state space}

\author{Richard D. J. G. Ho}
\email{richard.ho@ed.ac.uk}
\affiliation{
School of Physics,
University of Edinburgh, Edinburgh EH9 3JZ, United Kingdom}

\begin{abstract}
By analysing an n-dimensional generalisation of Thomas's cyclically symmetric attractor
we find that this chaotic dynamical system behaves like a random walk
constrained onto the surface of a hypersphere.
The growth of error is limited,
with qualitatively different behaviour depending on a control parameter.
For moderate values of the control parameter, linear growth of error is seen.
For low values of the control parameter, the error is limited by the
random walk behaviour.
Finally, we link this to the predictability of homogeneous isotropic
turbulence, which we find here also behaves like a constrained random walk.
\end{abstract}

\date{\today}

\maketitle

\section{Introduction}
\label{intro}

Understanding predictability allows us to measure the accuracy with which we can know
the future state of a system given knowledge of the current state.
Predictability is limited in chaotic systems, whereby initially close states
become exponentially different with time \cite{Lorenz1963}.
The difference between these two states is called the error.
Even within systems that become exponentially different, there may be
physical processes which allow a non-zero amount of predictability even
after a long period of time.
Real world systems often have large numbers of degrees of freedom
and are chaotic,
but also have non-zero predictability \cite{Lorenz1969}.

In this paper, we first look at 
a random walk constrained to 
the surface of an n-dimensional hypersphere.
Under this constraint, 
the difference between an initial state and its evolved state after a given period of
time can be predicted, with the theoretical prediction agreeing with simulation.
We then analyse a specific high dimensional chaotic system.
The difference
between an initial state and its evolved state
is consistent with 
a constrained random walk like behaviour.

Furthermore, when we analyse the
predictability of this chaotic system, we find that
the error does not grow exponentially for all times,
but instead has a limited growth rate.
The nature of this limited growth rate depends on the control parameter
with qualitatively different behaviour depending on the parameter.

Finally, we link the findings from this specific system and recent results from
simulations of hydrodynamic turbulence.
The results are qualitatively similar, suggesting a similar physical origin.
This further suggests that such random walk like behaviour may be a universal
feature of systems constrained by their energy.

\section{Random walk on a unit hypersphere}

We may consider a random walk
on the surface of hypersphere, centred on the origin, with $n$ dimensions.
If we take $\theta$ to be the angle between a point and its origin, then the
evolution of a point on the surface of this hypersphere obeys \cite{Caillol2004}
\begin{equation}
\cos (\theta) \sim \exp(-t/\tau) \ ,
\end{equation}
where $\tau$ is some diffusion rate and $t$ the time.
We take a point on the hypersphere to have a Euclidean norm $E$.
By the cosine rule, this means that
the difference, $D$, between a point and its origin grows as 
\begin{equation}
\label{eq:Dprediction}
D = 2E(1-\exp(-t/\tau)) \ .
\end{equation}
We test the prediction in Eqn.~(\ref{eq:Dprediction}) numerically.
We initialize an array of numbers, $x_i$, using a standard normal distribution,
which is the special case of a Gaussian distribution with zero mean and unit variance.
The hypersphere is of dimension $n$, with $i$ labelling the dimension and ranging from 1 to $n$.
We then normalize this array such that $E = \sum_i x_i^2 = 1$.

At each step, $t$, a random walk, $r_i(t)$, is generated.
This random walk assigns $+\sqrt{s/n}$ or $-\sqrt{s/n}$ to each dimension 
of the hypersphere
according to a Bernoulli distribution, where $+$ and $-$ are equally likely.
The distribution is updated at each step, with
\begin{equation}
r_i (t) = 
\begin{cases}
+\sqrt{s/n} \quad &\text{with probability} = 1/2 \\
-\sqrt{s/n} \quad &\text{with probability} = 1/2
\end{cases} \ .
\end{equation}
The normalisation by $\sqrt{s/n}$ ensures that the average step is of size $s$,
because the Euclidean norm of $r_i$, $|r|$, has $|r| = s$.
We then update $x_i(t + 1) = x_i(t) + r_i(t)$. After this update, we renormalize $x_i(t + 1)$
such that $E (t+1) = \sum_i (x_i(t+1))^2 = 1$.

The initial distribution is evolved for 1000 steps to further randomize the distribution.
The initial state of the system is defined as the state of the system after these 1000 steps.
It would be interesting to see how an arbitrary distribution would change in time under 
this set-up if it were not constrained. 
However, this is outside the scope of the current investigation.

After this process, we take the initial state of the system, evolve it, and
measure $D$, where
\begin{equation}
\label{eq:Ddefinition}
D(t) = \sum_i^n (x_i (t) - x_i(0))^2 \ ,
\end{equation}
and where $t = 0$ is defined as 
the step corresponding to the initial state.

\begin{figure}
\centering
\includegraphics[width=0.5\textwidth]{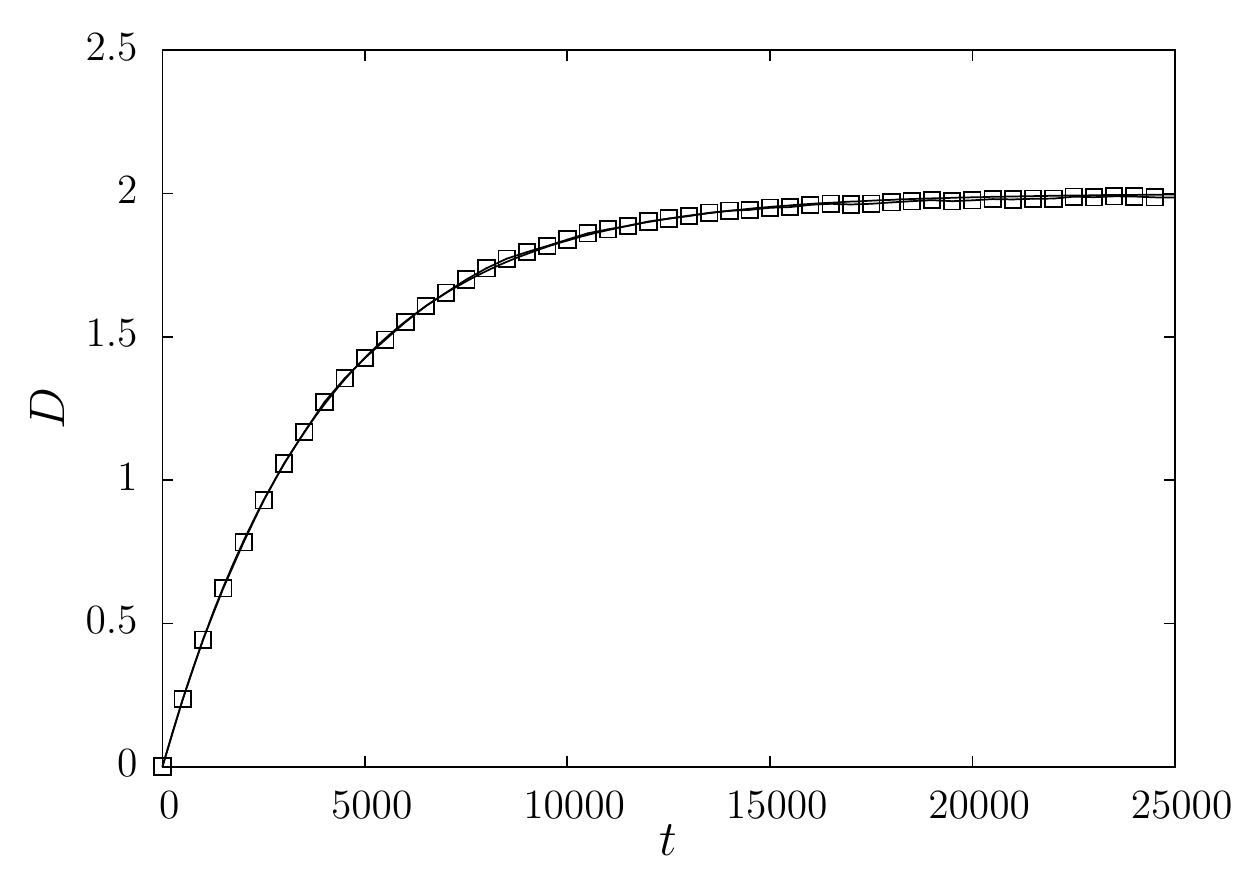}
\caption{Difference, $D$, as a function of step, $t$, for a random walk constrained on the surface of a unit hypersphere in $n$ = 100,000 dimensions, with step size, $s$ = 0.0005. Solid black line is prediction, squares are simulation results.}
\label{fig:crwfig}
\end{figure}

In Figure.~\ref{fig:crwfig} we show the result of a random walk constrained onto the 
surface of a unit hypersphere in $n$ = 100,000 dimensions. The step size, $s$, is 0.0005.
The solid black line is the prediction, whereas the squares are the simulation
results.
As can be seen from the figure, the predicted and measured results show very good agreement.
We find that the analogous quantity to the diffusion constant, $\tau$, 
has $1/\tau = s/2$.

\section{Application to an example chaotic system}

In this section, we try to relate the above results regarding random walks on
hyperspheres to many dimensional chaotic systems.
A three-dimensional dissipative cyclically symmetric dynamical system was proposed
by Thomas \cite{Thomas1999}
\begin{align}
\partial_t x(t) &= \sin(y(t)) - bx(t) \nonumber \\
\partial_t y(t) &= \sin(z(t)) - by(t) \\
\partial_t z(t) &= \sin(x(t)) - bz(t) \nonumber \ ,
\end{align}
where $\partial_t$ is the derivative with respect to time, $t$.
The control parameter $b$ represents a linear damping in the system,
which if low enough allows chaotic dynamics.

Thomas's cyclically symmetric attractor (TCSA) can be expanded to
$n$ dimensions as
\begin{equation}
\label{eq:ndimTCSA}
\partial_t x_i (t) = \sin(x_{i+1}(t)) - bx_i(t) \ ,
\end{equation}
where $i$ ranges from 1 to $n$, and $\partial x_n (t) = \sin(x_1(t)) - bx_n(t)$.
This system has been thoroughly examined and, at $b = 0$, shows behaviour like a 
random walk, with independent and identically distributed 
steps \cite{Sprott2007,Chlouverakis2007}.
In this system, energy, $E$, is defined $E(t) = \sum_i (x_i(t))^2$.
Such a system is interesting to compare to the random walk constrained on a hypersphere because,
for $b > 0$, $E$ varies only slightly.

The simulation results gathered here come from solving
TCSA in $n = 10,000$ dimensions using Euler's method.
Using prior tests, it was found that,
for high dimensions, 
an increase in dimension did not change the dynamics or results appreciably,
but did allow for smoother statistics.
Also, for $b \lesssim 0.01$, $E \sim 1/b$, but for higher values of
$b$, $E$ was higher than this. 
As such, the field is initialized with a Gaussian distribution with zero mean and 
standard deviation $1/\sqrt{2b}$.
This initialisation allows results which quickly reach a steady state of energy.
Even so, the simulation is evolved for 50 time units to remove any
artefacts from the initialisation.

After the initial evolution, both the difference, $D$, and error, $E_d$, are measured.
Difference $D$ is defined as in Eqn.~(\ref{eq:Ddefinition}).
To measure the evolution of the error, we create field $y_i(0) = x_i(0) + \delta_i$,
where $\delta_i$ is a small white noise perturbation.
Field $y_i(t)$ is then evolved according to Eqn.~(\ref{eq:ndimTCSA}),
with $x$ replaced by $y$.
The error is defined
\begin{equation}
E_d(t) = \sum_i (x_i (t) - y_i(t))^2 \ .
\end{equation}

\begin{figure}
\centering
\includegraphics[width=0.5\textwidth]{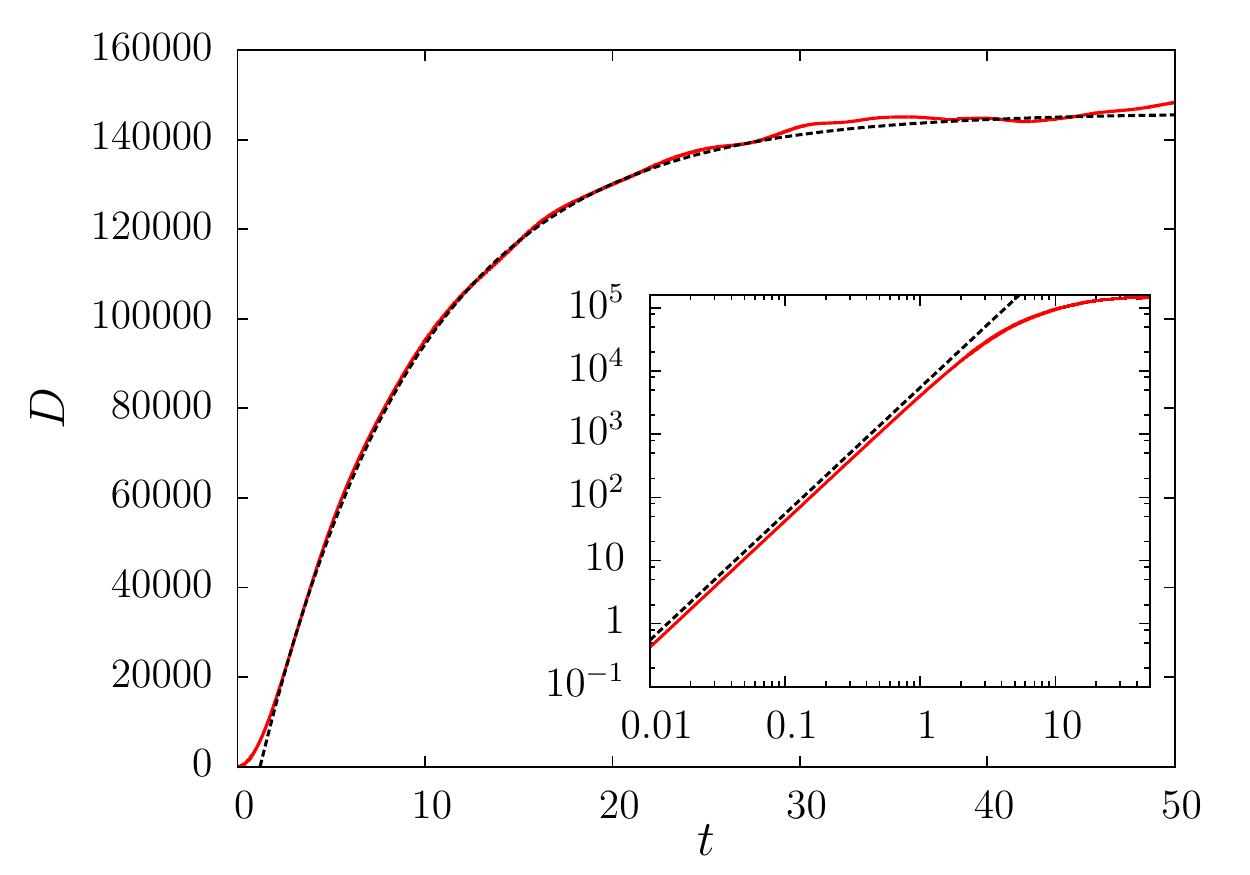}
\caption{Difference, $D$, as a function of time for a TCSA in $n$ = 10,000 dimensions, $b = 0.1$. Dashed black line is the expected behaviour of a random walk on a sphere, offset by a small time. Inset
is same as main plot, but log-log. It shows $t^2$ behaviour at small time.}
\label{fig:tccarw}
\end{figure}

Figure \ref{fig:tccarw} shows $D$ as a function of time for TCSA in $n$ = 10,000 dimensions
with $b = 0.1$.
For low $t$ there is an anomalous diffusion like behaviour where $D \sim t^2$,
as is seen in the inset of the figure.
However, for high $t$, $D$ behaves similar to the prediction for a random walk on a hypersphere
given in Eqn.~(\ref{eq:Dprediction}), but with a slight offset of time.
At small times, the behaviour is dominated by the chaotic dynamics
which produces the anomalous diffusion, but at large times, the system behaves
like a constrained random walk.

\begin{figure}
\centering
\includegraphics[width=0.5\textwidth]{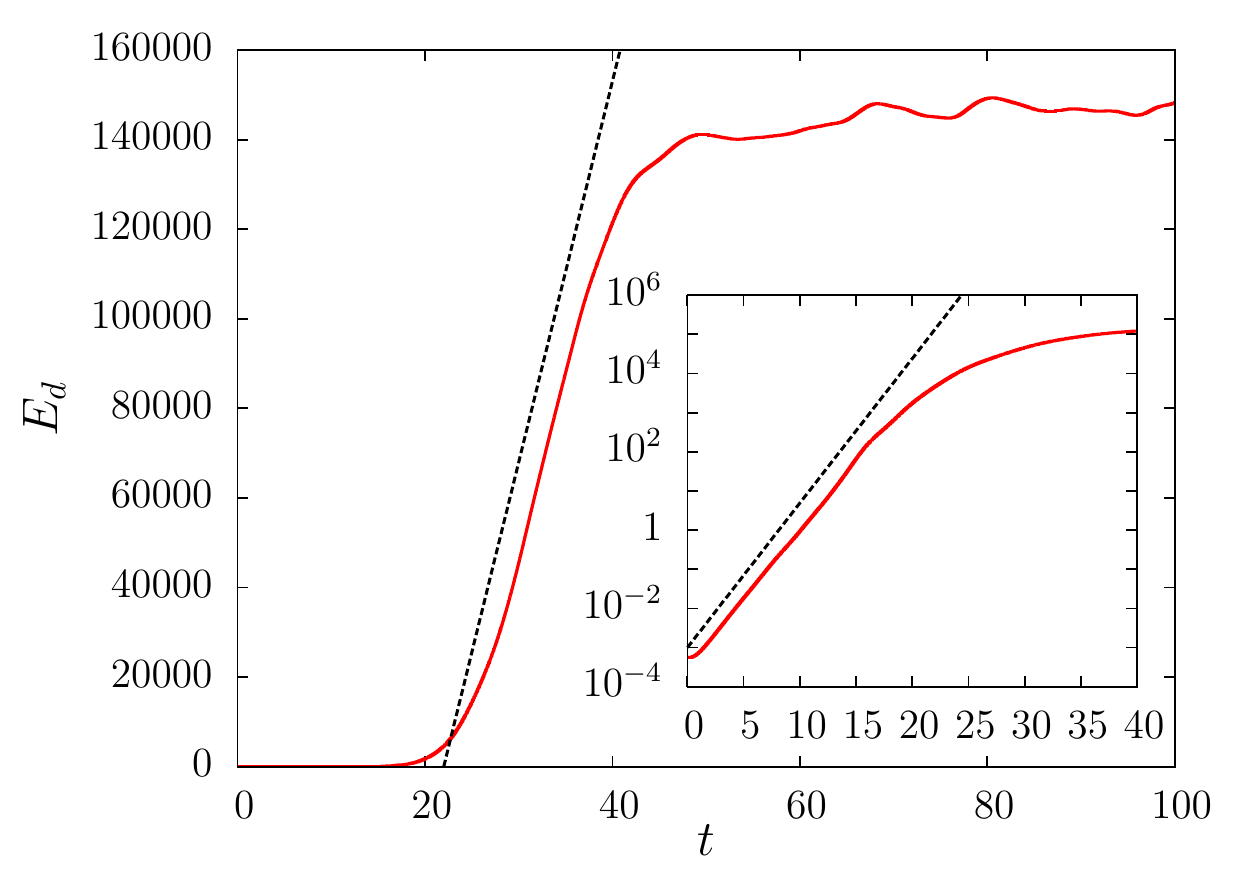}
\caption{Error, $E_d$, as a function of time for TCSA in $n$ = 10,000 dimensions, $b = 0.1$. Dashed black line is a linear behaviour with gradient $8500$, offset by a small time. Inset is same as main plot, but lin-log. The inset dashed line shows $exp(\lambda t)$ behaviour 
at small time, with $\lambda = 0.85$.}
\label{fig:highbchaos}
\end{figure}

Figure \ref{fig:highbchaos} shows $E_d$ as a function of time for TCSA in $n = 10,000$ dimensions
with $b = 0.1$.
This is the same system as shown in Fig.~\ref{fig:tccarw}.
For small time, error grows exponentially as expected in a chaotic system.
However, for large times there is a limit on growth such that $E_d$ starts to grow
linearly in time for an appreciable duration. 
Because the system has finite energy, saturation occurs at $2E$.

\begin{figure}
\centering
\includegraphics[width=0.5\textwidth]{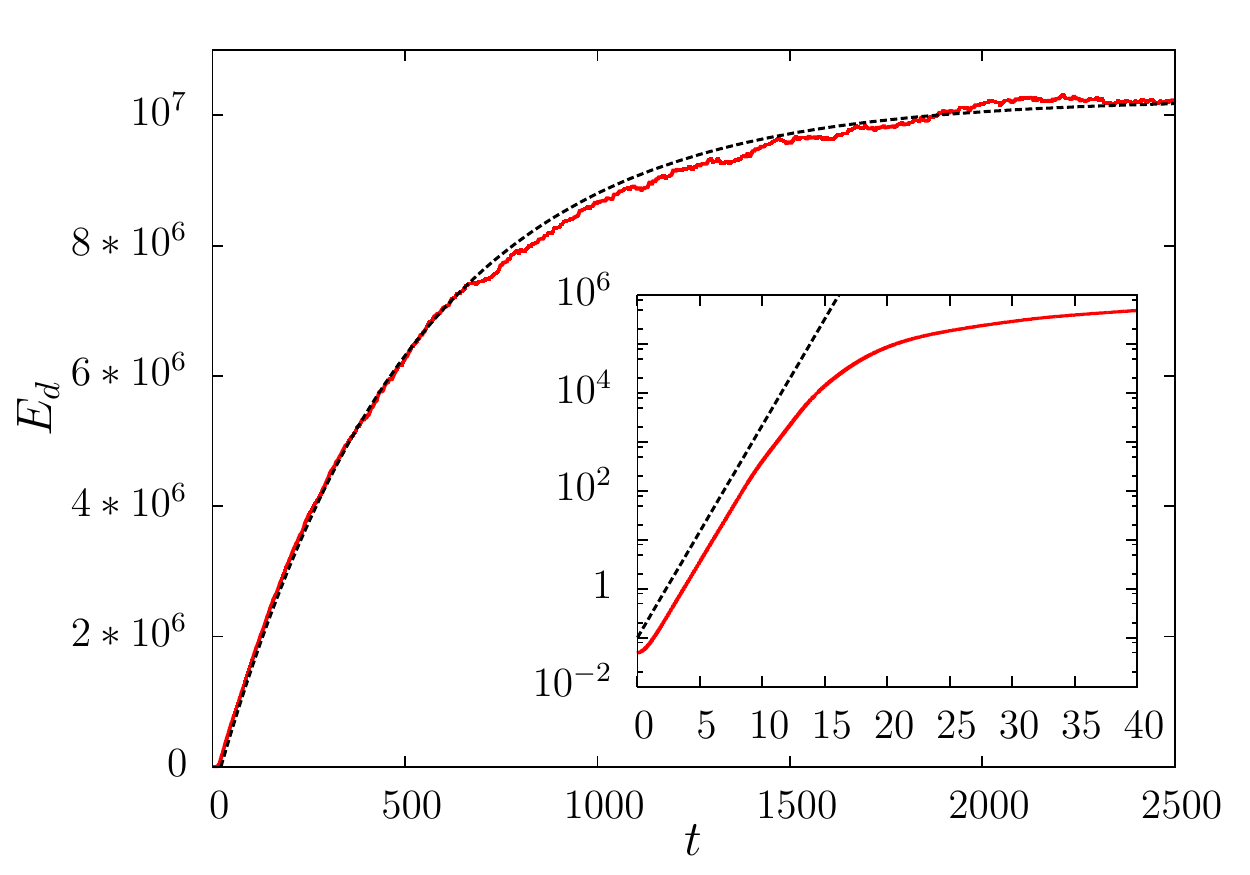}
\caption{Error, $E_d$, as a function of time for a TCSA in $n$ = 10,000 dimensions, $b = 0.001$. Dashed black line is the expected behaviour of a random walk on a sphere, offset by a small time. Inset is same as main plot, but lin-log. It shows $exp(\lambda t)$ behaviour at small time,
with $\lambda = 1$.}
\label{fig:lowbchaos}
\end{figure}

Surprisingly, for lower values of $b$, qualitatively different behaviour is found for $E_d$.
Figure \ref{fig:lowbchaos} shows $E_d$ as a function of time for TCSA in $n = 10,000$ dimensions
with $b = 0.001$.
Whilst there is still an initial exponential behaviour, with a similar Lyapunov exponent
to the case of $b = 0.1$, the long term behaviour is very different.
Instead of being linear, the long term behaviour of $E_d(t)$ looks like that of $D(t)$.

The growth of $D$ for $b = 0.001$ is slightly different than for $b = 0.1$.
For $b = 0.1$, the $t^2$ growth transitions smoothly into the late $2E(1-\exp(-t/\tau))$ growth.
However, for $b = 0.001$, there is a kink in the graph such that the $t^2$ part can
have a higher gradient than the $2E(1-\exp(-t/\tau))$ part.

The limiting behaviour of $E_d$ for low $b$ is presumably due to the following physical effect.
We may consider an initial point in state space, after time $t$, the evolved state is
distance $d$ from the initial state.
If we have two points which are initially close to the initial state, 
then after time $t$, they should
be at most distance $2d$ apart.
As such, if $D$ has a certain time dependence, we should expect $E_d$ to be bounded
in time by a similar time dependence.
Because $E_d$ grows qualitatively differently for low and high $b$, 
the high $b$ behaviour of $E_d(t)$ must be limited by a different effect.

To investigate the different behaviour for different $b$, 
$b$ is varied from 0.001 to 0.22.
We measure the maximum gradient for the growth of $E_d$ and $D$, 
which are labelled $m_E$ and $m_D$ respectively.
An average is taken of 10 runs each value of $b$.

\begin{figure}
\centering
\includegraphics[width=0.5\textwidth]{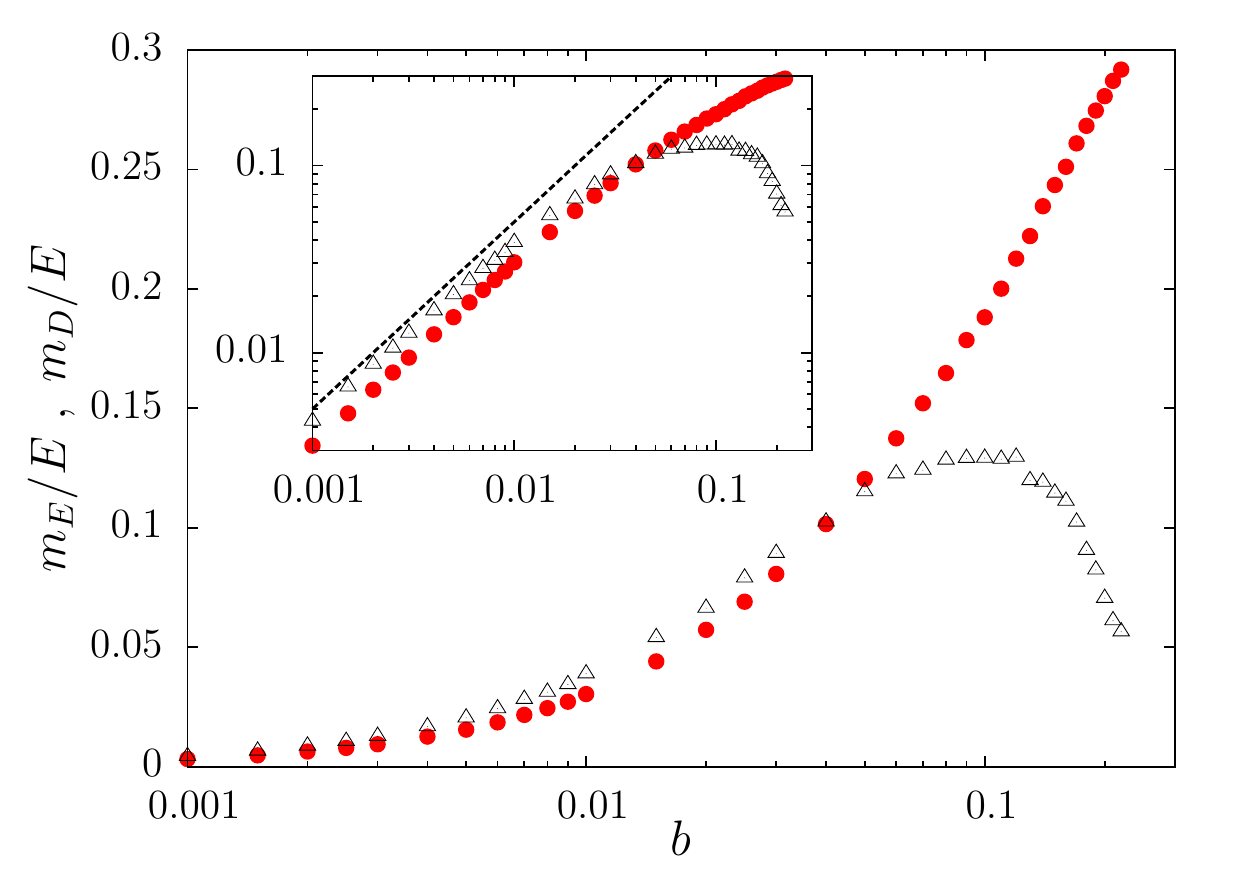}
\caption{Maximum gradients $m_E$ (triangles) and $m_D$ (circles) 
normalized by $E$ for different $b$.
Inset is same as main plot but log-log, inset dashed line shows a linear relation.}
\label{fig:ratevsb}
\end{figure}

Figure \ref{fig:ratevsb} shows $m_E$ (triangles) and $m_D$ (circles) 
normalized by energy $E$ for varying $b$.
From the figure we see that, going from low to high $b$, $m_E$ and $m_D$ increase
proportionally to each other. Physically, an increase in $m_D$ allows $m_E$
to also increase because the bound on the growth of $m_E$ is increased.
The values for $m_E$ are slightly higher than $m_D$,
but this is possible since the bound on $E_d$ is $2D$.
For $b$ = 0.001, the ratio $m_E / m_D = 1.35 \pm 0.04$.

However, around $b \approx 0.05$, an increase in $m_D$ is no longer accompanied
by an increase in $m_E$. In this region, a different effect must be limiting the 
growth of $E_d$, 
Around $b = 0.1$, the rate of $E_d$ growth is independent of $b$,
and $m_E$ is roughly proportional to $E$,
because $E$ is almost inversely dependent on $b$ in this region.

\section{Comparison to turbulence}

We compare the qualitative behaviours just described in the n-dimensional TCSA
with other high dimensional chaotic systems to see if there are analogous behaviours.
One such physical system is that of turbulent hydrodynamics.
Turbulent flows are characterized by their Reynolds number, Re = $uL/\nu$,
where $u$ is the typical velocity, $L$ a typical length scale, and $\nu$ the kinematic viscosity.
A fluid field $\bm{u}$, evolving according to the Navier-Stokes equation, with high enough Re
becomes turbulent, which means that the flow develops highly complex spatial
and temporal structure
and becomes chaotic.

Many recent exciting directions in understanding turbulence have come from the 
perspective of dynamical systems theory.
These include relaminarisation of turbulence
in pipe flow \cite{eckhardt2007turbulence} and directed percolation models for
understanding turbulent lifetimes \cite{lemoult2016directed}.
In this section we try to use the insights gained in the above section and apply
them to the case of turbulence.
We compare the behaviour of $D$ and $E_d$ between the n-dimensional
TCSA and hydrodynamic turbulence.
In hydrodynamics the difference, $D$, and error, $E_d$, are defined
\begin{align}
D (t) &= \frac{1}{2} \int (\bm{u} (\bm{x},t) - \bm{u} (\bm{x},0))^2 \ dV \ , \\
E_d(t) &= \frac{1}{2} \int (\bm{u}_1 (\bm{x},t) - \bm{u}_2 (\bm{x},t))^2 \ dV \ ,
\end{align}
which are integrals over all space.

\begin{figure}
\centering
\includegraphics[width=0.5\textwidth]{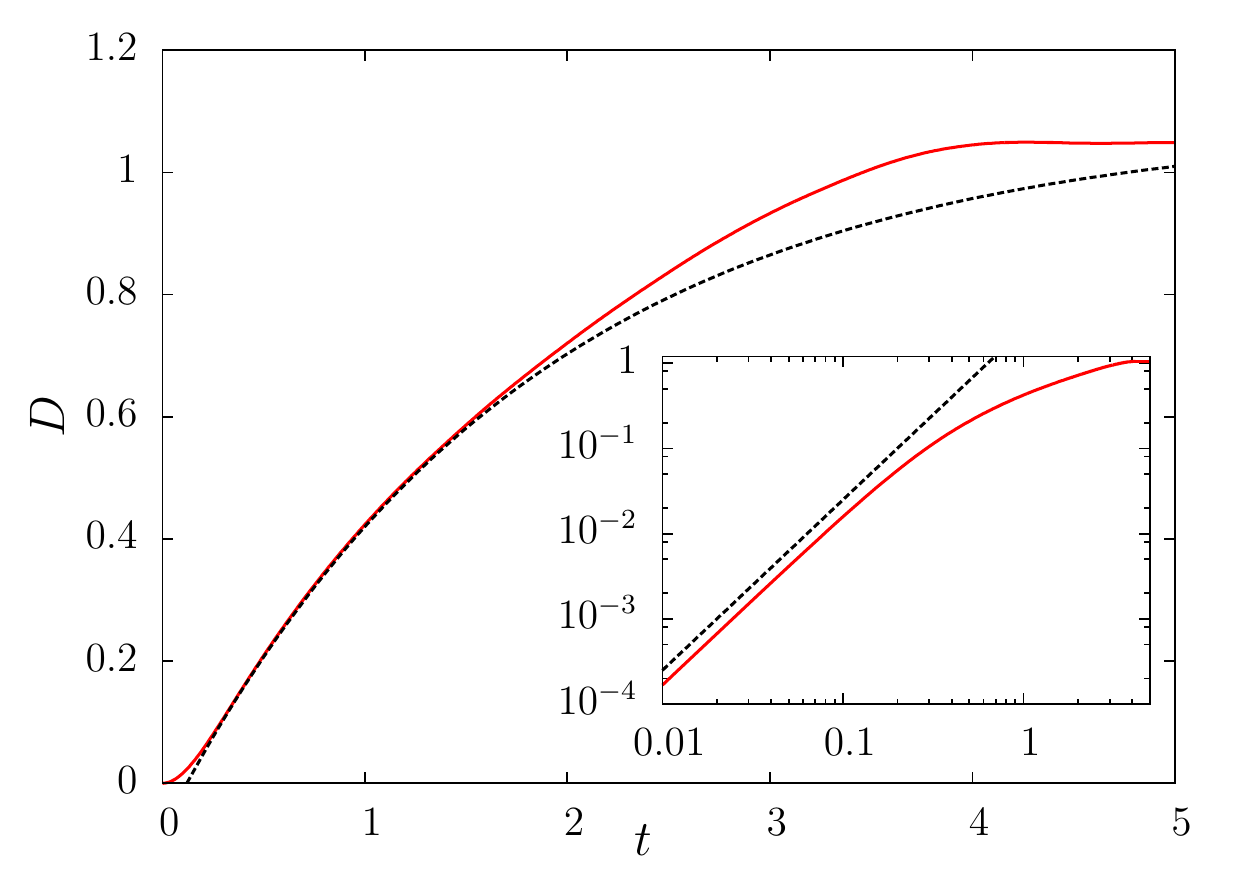}
\caption{Difference, $D$, as a function of time, in simulation units, for Re = 490.
Red (grey) solid line is measured $D$ and dashed black line is a comparison to Eqn.~(\ref{eq:Dprediction}). Inset is same as main plot but log-log. Inset dashed line
shows $t^2$ behaviour.}
\label{fig:randomwalkinphasespace}
\end{figure}

Figure \ref{fig:randomwalkinphasespace} shows the result of measuring $D$ for
a simulation of homogeneous isotropic turbulence with Re = 490.
The details of this simulation are the same as those used for the main results 
in \cite{Berera2018}.
As can be seen from the figure, the growth of $D$ is well approximated as following
the prediction in Eqn.~(\ref{eq:Dprediction}).
The departure from prediction at later time is due to the fact that
the assumption of constant energy is not as good in turbulence
as it is for the n-dimensional TCSA.
As well, similar to the n-dimensional TCSA, at small time there is an anomalous diffusion
type growth where $D \sim t^2$, which can be seen in the inset.
As such, we can say that the behaviour of hydrodynamic turbulence is consistent
with that of a random walk on a high dimensional hypersphere.

\section{Conclusion and discussion}

For three-dimensional turbulence with moderate dissipation, 
it has recently been seen that $E_d$ grows
linearly in time with a rate proportional to the dissipation \cite{Berera2018,Boffetta2017}.
This is analogous to the behaviour of the n-dimensional TCSA found in this paper 
at moderate $b$.
However, it has also been found that for very low dissipation there is a qualitatively
different type of behaviour for hydrodynamic turbulence. 
In two-dimensional flows, which allow for lower dissipation than three-dimensional ones,
the evolution of $E_d(t)$ is more similar to that of $D(t)$
in Fig.~\ref{fig:randomwalkinphasespace}.
An example of this can be seen in \cite{Boffetta2001}.
This suggests that the same bounding of error growth by $D$ that has been found for the
n-dimensional TCSA here, also may hold in other high dimensional chaotic systems.
Thus, in turbulence, we see both type of bounding effect seen here
for the n-dimensional TCSA, linear and random walk like, as well as a dependence
on a dissipation/damping parameter.


The fact that there are analogous behaviours
for both $D$ and $E_d$ between the n-dimensional TCSA and
hydrodynamic turbulence suggests that this may represent some more universal
behaviour of high dimensional chaotic systems which are bounded by some energy constraint.
We note finally that $D$ and $E_d$ are important quantities from the point of
studies of predictability. They each give an idea of the way in which bounds
may be made on future prediction given the degree of current knowledge of a system.
There are many systems that are bounded by finite energy, but are otherwise chaotic,
and the limit of their predictability should be that of a random walk on the
surface of a hypersphere.

\begin{acknowledgments}
This work has used resources from ARCHER \cite{archer} via the Director's Time budget.
R.D.J.G.H is supported by the U.K. Engineering and Physical Sciences Research Council (EP/M506515/1).
The author thanks Arjun Berera for helpful discussion.
\end{acknowledgments}

\end{document}